\documentclass[graybox]{article}

\usepackage{mathptmx}       
\usepackage{helvet}         
\usepackage{courier}        
\usepackage{type1cm}        
\usepackage{makeidx}         
\usepackage{graphicx}        
\usepackage{multicol}        
\usepackage[bottom]{footmisc}

\usepackage{tabularx}

\makeindex             

\begin{document}

\title{Second-Order Assortative Mixing\\  in Social Networks}

\author{Shi Zhou$^*$, ~Ingemar J.~Cox   \\
	 Department of Computer Science\\ University College London (UCL), UK \\
	  $^*$Email: {s.zhou@ucl.ac.uk} \\
	\and 
	Lars K.~Hansen  \\
	Department of Applied Mathematics and Computer Science\\ Danish Technical University (DTU), Denmark  
}

\date{}

%

\maketitle

\abstract{
In a social network, the number of links of a node, or node degree, is often assumed as a proxy for the node's importance or prominence within the network. 
It is known that social networks exhibit the (first-order) assortative mixing, i.e.~if two nodes are connected, they tend to have similar node degrees, suggesting that people tend to mix with those of comparable prominence. 
In this paper, we report the {\em second-order} assortative mixing in social networks. 
If two nodes are connected, we measure the degree correlation between their  most prominent  {\em neighbours}, rather than  between the two nodes themselves. 
We observe very strong second-order assortative mixing in social networks, often significantly stronger than the first-order assortative mixing. 
This suggests that  if two people interact in a social network, then the importance of the most prominent person each knows is very likely to be the same.
This is also true if we measure the average prominence of neighbours of the two people.  
This property is weaker or negative in non-social networks.
We investigate a number of possible explanations for this property. However, none of them was found to provide an adequate explanation.  We therefore conclude that second-order assortative mixing is a new property of social networks.
}


\section{Background}
A network or graph consists of nodes  connected together via links. Networks
are utilised in many disciplines. The nodes model physical elements such as people, proteins or
cities, and the links between nodes represent connections between them, such as contacts,
biochemical interactions, and roads. 
%
%
In recent years studying
the structure, function and evolution of networked systems in society and nature has become a major
research focus~\cite{wasserman94,watts99,barabasi02,bornholdt02,dorogovtsev03a,pastor04}.



The  degree, $k$, of a node is defined as the number of links the node possesses.
The probability distribution of node degrees is indicative of a network's global connectivity. For
example random graphs with a Poisson degree distribution~\cite{erdos59} have most nodes with
degrees close to the average degree.  In contrast, many complex networks in nature and society are scale-free graphs~\cite{Barabasi99} exhibiting a power-law degree distribution, where many
nodes have only a few links and a small number of nodes have very large numbers of links. However,
the degree distribution alone does not provide a full description of a network's topology. Networks
with exactly the same degree distribution can possess other
properties that are vastly different~\cite{Maslov04,mahadevan06,zhou07b}.

One such property, is the mixing pattern between the two end nodes of a link~\cite{Pastor01,newman02},
i.e.~the joint probability distribution of a node  with degree $k$ being connected to a node with degree
$k'$. In general, biological and technological networks are {\emph{disassortative}} mixing meaning that
well-connected nodes tend to link with poorly-connected nodes, and vice versa. In contrast, social networks,
such as collaborations between film actors or scientists, exhibit {\em assortative} mixing, where nodes with
similar degrees tend to be connected.

To quantify this mixing property, Newman~\cite{newman02} proposed the assortative coefficient,~$r$, where
$-1\leq r\leq 1$.
%
%
It is derived by considering the Pearson correlation between two sequences, where corresponding elements in
the two sequences represent the degree of the nodes at either end of a link in the network. For a directed network, the degree of the starting node of a link is contained in one sequence, and
the degree of the ending node is in the other sequence. The number of elements in each sequence is the number
of links. For an undirected network, as all the networks studied in this paper, each undirected link is
replaced by two directed links pointing at opposite directions. Thus the number of elements in a sequence
is twice the number of links.
 
A network with assortative mixing is characterised by a possible value of $r$; where  
$r=1$ corresponds to a perfect assortative mixing, i.e., every link connects two
nodes with the same degree. 
A network with disassortative mixing has a negative value of $r$; where  
$r=-1$ corresponds to a perfect disassortative mixing, i.e., every link connects two
nodes with difference degrees. 
When $r$ equals or close to 0, there is no  degree correlation, i.e., the network
is random or neutral in terms of degree mixing. 

The mixing pattern
has been studied as a fundamental property of networks, and the assortative coefficient $r$ has been widely
used to measure this property.

\section{Second-Order  Mixing Pattern}

We now introduce and define a related property which we refer to as the {\em second-order} mixing pattern.

\subsection{Definition of ${\cal{R}}_{max}$ and ${\cal{R}}_{avg}$}

Following Newman's definition of the
(first-order) assortative coefficient $r$~\cite{newman02},    we
define ${\cal{R}}_{max}$ as the second-order assortative coefficient   based on the neighbours maximum degree,  
\begin{equation}
{\cal{R}}_{max}=\frac{\displaystyle L^{-1}\sum_{ {i} } K_i K'_i - \left[ {1\over 2}L^{-1}\sum_{
{i} }(K_i+K'_i)\right]^2} {\displaystyle {1\over2} L^{-1}\sum_{ {i} }({K_i}^2+{K'_i}^{\,2})-\left[
{1\over2}L^{-1}\sum_{ {i} }(K_i+ K'_i) \right]^2 }, \label{eq:SecondOrder}
\end{equation}
where $L$ is the number of links in the network, $K_i$ and $K'_i$ are the neighbours maximum  degrees of the two nodes $a$ and $b$
connected by the link $i$, i.e.~$K_i = \max(k_{n_a}: n_a \in N_{a\setminus b})$ and $K'_i = \max(k_{n_b}: n_b \in
N_{b\setminus a})$; $N_{a\setminus b}$ denotes the
set of neighbours of node~$a$, excluding node $b$; and $N_{b\setminus a}$ denotes the
set of neighbours of node~$b$, excluding node $a$. 

\begin{figure}
\centering
	\includegraphics[width=4cm]{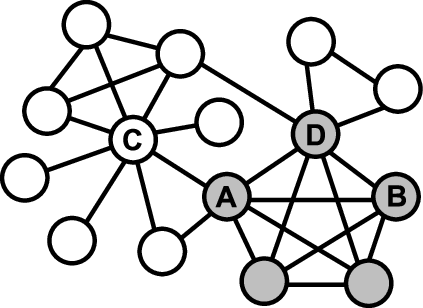}
	\caption{ Examples of node excess degrees, which is node degree  minus
		one. Consider the link between nodes $A$ and $B$, the excess degree of node $A$ is 5; and the
		neighbours maximum excess degree of node $A$ is 7, which is the excess degree of node $C$. }
	\label{fig:example}
\end{figure}

Note that when calculating the first and second assortative coefficients, we actually use the \emph{excess} degree~\cite{newman02}, which is degree minus
one. 
This is because, when considering two connected
nodes, $A$ and $B$,  the  neighbourhood of $A$ is defined to exclude
$B$. And likewise for the neighbourhood of $B$.
See Fig.~\ref{fig:example} for examples.

Similarly we define ${\cal{R}}_{avg}$  as the second-order assortative coefficient based on
the neighbours average degrees by replacing $K_i$ and $K'_i$ in the above equation as $K_i =
\frac{1}{k_a}\sum_{n_a \in N_{a\setminus b}}{k_{n_a}}$ and $K'_i = \frac{1}{k_b}\sum_{n_b \in N_{b\setminus
a}}{k_{n_b}}$. 

\subsection{Results}

\begin{table}
	
	\centering
\caption{ {Properties of the networks under study}. 
	Properties shown are the numbers of nodes and links; the assortative coefficients $r$,
	${\cal{R}}_{avg}$ and ${\cal{R}}_{max}$ with the corresponding expected standard deviation $\sigma_r$, $\sigma_{avg}$ and $\sigma_{max}$; and the
	average clustering coefficient of nodes in a network,\,$\langle C\rangle$.
({\bf a})~Film actor
collaborations~\cite{Barabasi99}, where two actors are connected if they have co-starred in a film;
({\bf b})~Scientist collaborations~\cite{newman01a}, where two scientists are connected if they have
co-authored a paper in condense matter physics;
({\bf c})~Jazz musician network~\cite{gleiser03}, where two musicians are connected if they have played in a band;
({\bf d})~Secure email network~\cite{boguna04}, where a link represent a secure email exchange between two trusted users using the
Pretty Good Privacy (PGP) algorithm;
({\bf e})~General email network~\cite{guimera03},  where email exchanges take place at a university, including a large amount of unsolicited emails;
({\bf f})~Western States Power Grid of the United States~\cite{watts98};
({\bf g})~\emph{C.\,elegans} metabolic network~\cite{jeong00}, where two metabolites are
connected if they participate in a biochemical reaction;
({\bf h})~the protein interactions of the yeast~\emph{Saccharomyces
cerevisiae}~\cite{Maslov02a,colizza05};
({\bf i})~Internet~\cite{Pastor01} (http://www.routeviews.org/), where two service providers are
connected if they have a commercial agreement to exchange data traffic;
({\bf j})~the random graphs~\cite{erdos59};
and ({\bf k})~the Barab\'asi-Albert (BA) graphs~\cite{Barabasi99}.
}\label{table:networks}
\vspace{2mm}
\renewcommand{\arraystretch}{1.2}
\renewcommand{\tabcolsep}{0.33pc}
\begin{tabular}{lrrrcrcrrr}
\hline\noalign{\smallskip}
Network  &   Nodes  &Links    &   $r$ & ${\sigma}_r$  & ${\cal{R}}_{avg}$ & ${\sigma}_{avg}$
&   ${\cal{R}}_{max}$ & ${\sigma}_{max}$  &  $\langle C\rangle$ \\
\noalign{\smallskip}\hline\noalign{\smallskip}
(a) Film actor   &   82,593    &   3,666,738    &   0.206& 0.013  &   0.836& 0.027                &   0.813& 0.009     &   0.75 \\
(b) Scientist   &   12,722   &   39,967 &   0.161& 0.007   &   0.680& 0.014   &   0.647& 0.005    &   0.65   \\
(c) Musician  &   198    &   2,742  &   0.020& 0.019   &   0.543& 0.023   &   0.307&  0.029   &   0.62   \\
(d) Secure  email     &   10,680  &   24,316    &   0.238& 0.007  &   0.653& 0.009   &   0.680&  0.007    &   0.27  \\
\noalign{\smallskip}\hline\noalign{\smallskip}
(e) General email     &   1,133  &   5,451       &   0.078& 0.014   &   0.242& 0.014   &   0.247&  0.014    &   0.22  \\
(f) Power grid &   4,941   &   6,594  &   0.004& 0.014   &   0.205& 0.015   &   0.258&  0.016    & 0.08  \\
(g) Metabolism    &   453   &   2,025    &   -0.226& 0.011   &   0.265& 0.032   &   0.263&  0.023    &   0.65 \\
(h) Protein      &   4,626   &   14,801     &   -0.137& 0.008   &   -0.046& 0.007   &   0.033&  0.009    &   0.09 \\
(i) Internet &   11,174   &   23,409   &   -0.195& 0.001   &   -0.097& 0.004   &   0.036&  0.008    &   0.30    \\
(j) Random graph    &   10,000  &30,000    &   $\simeq0$& 0.009  &    $\simeq0$& 0.011   &    $\simeq0$& 0.006   &   $\simeq0$   \\
(k) BA graph   &     10,000  &30,000  &    $\simeq0$& 0.004   &    $\simeq0$& 0.008   &    $\simeq0$& 0.008    &   $\simeq0$ \\
\noalign{\smallskip}\hline\noalign{\smallskip}
\end{tabular}
\end{table}


We consider eleven networks, including five social networks, two biological
networks, two technology networks, and two synthetic networks based on random connections~\cite{erdos59} and
the Barab\'asi and Albert~\cite{Barabasi99} model, respectively.
Values of both the first-order and the second-order assortative coefficients, $r$,  ${\cal{R}}_{avg}$ and ${\cal{R}}_{max}$ are provided in
Table\,\ref{table:networks}.  

\subsubsection{Statistical significance}

The expected standard deviation $\sigma$ on the value of assortative coefficient $r$ can be obtained by the
jackknife method~\cite{efron79} as $\sigma^2=\sum_{i=1}^{L}(r_i-r)^2$, where $r_i$ is the value of $r$ for
the network in which the $i$-th link is removed and $i=1,2,...L$. And likewise for second-order assortative
coefficients ${\cal R}_{max}$ and ${\cal R}_{avg}$. For all cases shown in Table~1, the value of $\sigma$ is
very small ($<0.03$), which validates the statistical significance of the coefficients.

\subsubsection{Null hypothesis test}

A high correlation score between two value sequences must be tested against the null hypothesis. For each
network and each coefficient in Table~\ref{table:networks}, we randomly permuted the order of  degree values in one of the two degree sequences and
re-computed the coefficient. This was repeated 100 times and then we calculated the mean and standard
deviation. Our calculation shows that for each
network and each coefficient the mean value is close
to zero and the standard deviation is small. This result again confirms the  statistical significance of the
first and second-order assortative coefficients.

\subsubsection{Social networks}

Four social networks (a)-(d) show  positive values of first-order assortative coefficient, and notably they show significantly higher values of second-order assortative coefficients.  
This indicates that in these social networks, people judge on other individual's social status based on not only the individual's own prominence (e.g. the number of co-starred films or co-authored    publications), but more crucially, the prominence of is collaborators.   

Interestingly,    the \texttt{Musician} network  exhibits very low first-order assortative mixing although it shows one of the strongest second-order assortative mixing. In other words, although 
musicians in this network do exhibit  a strong social parity, it cannot be  revealed  by  measuring 
the prominence of musicians themselves; instead we must measure the prominence of other musicians that each musician has ever performed with. 
 
The {\tt Secure email} network's
strong second-order assortative mixing is  due to the security feather of this network where a person's security credit relies on endorsement from its contacts - the more credit a contact already has the more valuable its endorsement.

\subsubsection{Non-social networks} 

Other forms of networks  do not exhibit the very strong second-order correlations exhibited by social
networks. 

The {\tt General email} network is not considered as a typical social network, because it contains a large amount of unsolicited, one-way communications, such as notices and advertisements  forwarded from departmental secretaries to all students. Not surprisingly, this network's second assortative mixing is as weak as  the {\tt Metabolism} and {\tt Power grid} networks. 

The {\tt
  Internet} and {\tt Protein} networks, the second order assortative
coefficients are either zero ($R_{max}$) or negative ($R_{avg}$). 

As
expected, neutral random networks generated by   graph models are
completely uncorrelated, i.e. $R_{max}=R_{avg}=0$.

\subsection{Frequency distributions of links   }

\begin{figure}
	\centering
\includegraphics[width=15cm]{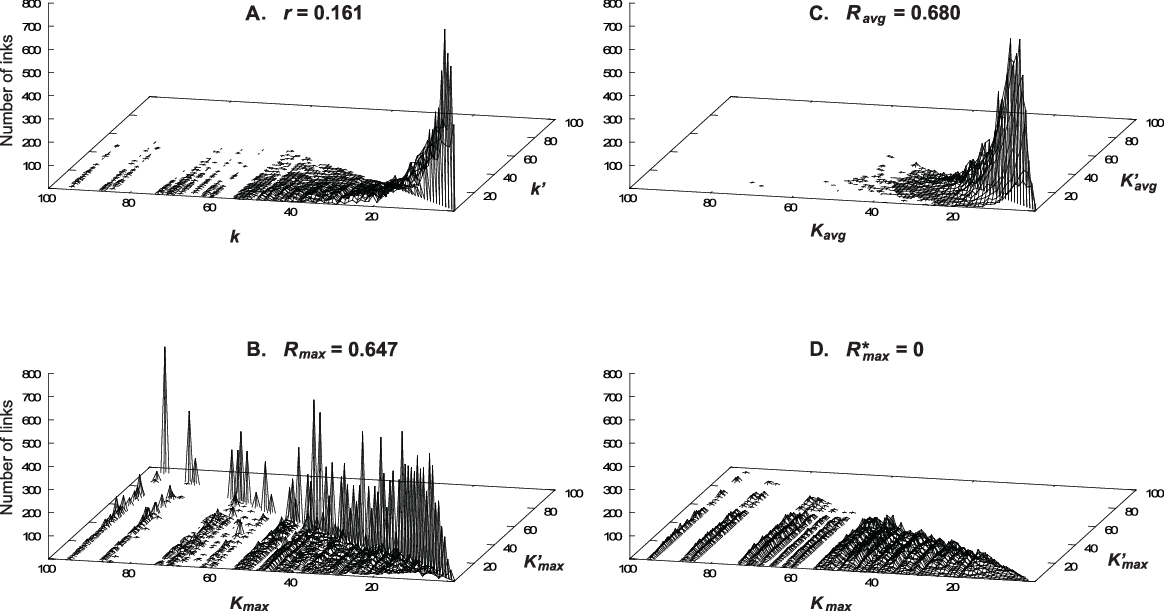}
\caption{ {The first and second-order assortative mixing in the \texttt{Scientist} network}. We show the
link frequency distribution as functions of ({A})~degrees $k$ and $k'$ of the two end nodes of a link,
with $k\geq k'$; ({B})~the neighbours maximum degrees of the two nodes, $K_{max}$ and $K'_{max}$, with
$K_{max}\geq  K'_{max}$; and ({C})~the neighbours average degrees, $K_{avg}$ and $K'_{avg}$, with
$K_{avg}\geq  K'_{avg}$, respectively. ({D})~is the same as ({B}), where links are randomly rewired while  preserving
the degree distribution. The maximum degree of the network is 97.} \label{fig:3D}
\end{figure}

Fig.~\ref{fig:3D} provides a more detailed look into the assortative mixing in   the {\tt Scientist} network. Fig.~\ref{fig:3D}(A) shows  there is a strong first-order correlation between node degrees when $k<20$, and  the correlation rapidly decreases with increasing degree, as expected for a scale-free
network.

For the second-order assortative mixing, Fig.~\ref{fig:3D}(B) shows a
very strong correlation for almost all values of the neighbours maximum degree $K_{max}$, where the link distribution along the diagonal
does not decrease with the increase of $K_{max}$.   
Of
course the correlation in Fig.~\ref{fig:3D}(B) is not perfect, and a second process appears to be uniform
noise. The noise might be better modelled as Gaussian which is probably due to the summation of many
nodes and the central limit theorem. If the neighbours average degree rather than the maximum is considered, we still observe a strong
correlation in Fig.~\ref{fig:3D}(C).

\section{Seeking  Possible Explanations}

Here we examine whether the second-order mixing is a \emph{new} topological property, i.e. whether it can be explained by other known properties of the networks.

\subsection{Increased Neighbourhood}


One may wonder whether the strong correlation scores associated with second-order assortative
mixing could simply be due to the increased  neighbourhood (from distance of one hop to two hops), as a node always has more second-order neighbours
than first-order neighbours. To exclude this possibility we also examined the $X$th-order assortative
coefficients, ${\cal R}_{max}$ and ${\cal R}_{avg}$, which are calculated using the maximum or average degree
within the neighbourhood of up to $X$ hops from each end node of a
link.  
Of course, if the
neighbourhood continues to increase, we observed that eventually the coefficients would increase and approach to one. This is to
be expected since eventually, the neighbourhood encompasses the entire network.
 
However, we observed that for {\em all} networks
under study, the values of the third-order coefficients were actually  smaller than the 2nd-order coefficients. 
This
suggest that the second-order assortative mixing cannot be explained by increased
neighbourhood. 

The fact that the third-order coefficients are smaller than the second-order coefficients has rich meanings. For technology networks, consider the Internet, where a network service provider only cares about the prominence of a customer (disassortative first-order mixing),   it does not know and  care   about who else the customer has linked with (neutral second-order mixing),  and care even less about those one step further away. For social networks,  one tends to match its collaborator's prominence (first-order assortative mixing) and the prominence of the collaborator's contacts (stronger second-order assortative mixing), but it does not know or care about contacts of the collaborator's contacts whom the collaborator does not know directly.  In other words, the value of social prominence vanishes rapidly after the second-order.

\subsection{High-Degree Nodes}

Another possible explanation for the high values of second-order
assortative coefficients considered, is that there are a few hub
nodes that are extremely well connected and dominate the network structure. To test this we removed the
best-connected node (together with the links attaching to it and any resulting isolated nodes) from the
networks and re-computed the coefficients. We also calculate the coefficients after removing the top 5
best-connected nodes. Results show that in all cases, the coefficients
change very little. For some networks, such as the \texttt{Secure email}, \texttt{Musician} and
\texttt{Metabolism} networks, the second-order coefficients became
stronger after the best-connected nodes
are removed.
This
suggest that the second-order assortative mixing cannot be explained by the existent of high-degree nodes.

\subsection{Power-Law Degree Distribution}

While high degree nodes do not explain the high second order
assortative mixing scores, the underlying heterogeneous power-law
structure of the networks was also a possible explanation. To exclude this possibility we used the random link rewiring
algorithm~\cite{Maslov04,zhou07b} to produce surrogate networks by randomly rewiring links while preserving the exact degree distribution of
the networks under study. 

Fig.~\ref{fig:3D}(D) illustrates the distribution of links as a function of $K_{max}$ and $K'_{max}$ in a
randomly rewired version of the \texttt{Scientist} network. The second-order assortative mixing in the original
network disappears completely in the randomised case.. 

This result shows that the second-order mixing is {\em not} determined by a network's degree
distribution, because two networks (the original and the randomised case) with the identical degree distribution show hugely different mixing patterns,
both in the first-order~\cite{Maslov04,zhou07b} and in the
second-order (see Fig.~\ref{fig:3D}).  

This result again demonstrates the limitation of
characterising network topology by degree distribution alone, and highlights the critical importance of characterising a network's topology using multiple properties from different aspects.

\subsection{Clustering Coefficient}

\begin{figure}
\centering
\includegraphics[width=10cm]{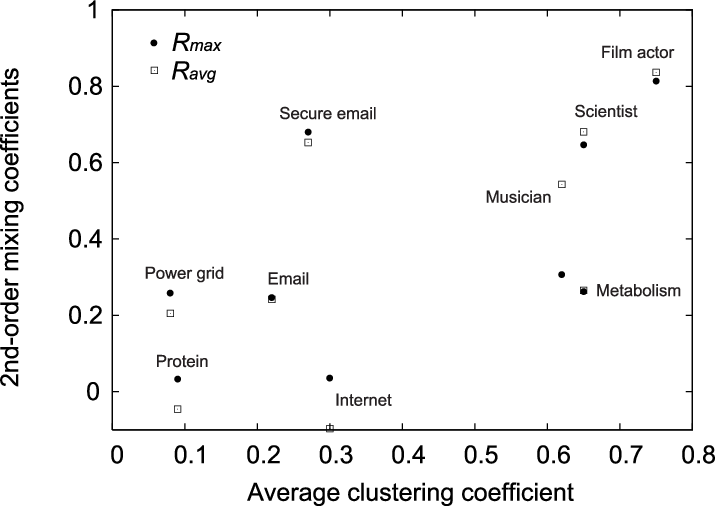}
\caption{ Second-order mixing coefficients vs
average clustering coefficient} \label{fig:C-R}
\end{figure}


We also examined whether the second-order assortative mixing is a consequence of the clustering behaviour
observed in many social networks, where one's friends are also friends of each other. This is quantified by
the clustering coefficient, $C_i$, which is defined as $C_i =\frac{e_i}{k_i(k_i-1)/2}$, where $k_i$ is the
degree of node $i$ and $e_i$ is the  number of connections between the node's neighbours~\cite{watts99}.
The average clustering coefficient, $\langle C\rangle$, is the arithmetic average over all nodes in the
network. 
Comparison of $\langle C\rangle$ against ${\cal{R}}_{avg}$ and ${\cal{R}}_{max}$ in
Table\,\ref{table:networks} and Fig.\,\ref{fig:C-R} shows that high values of the second-order 
coefficients occur for both high and low values of clustering coefficient. There is no correlation between them. 

\begin{figure}
\centering
\includegraphics[width=10cm]{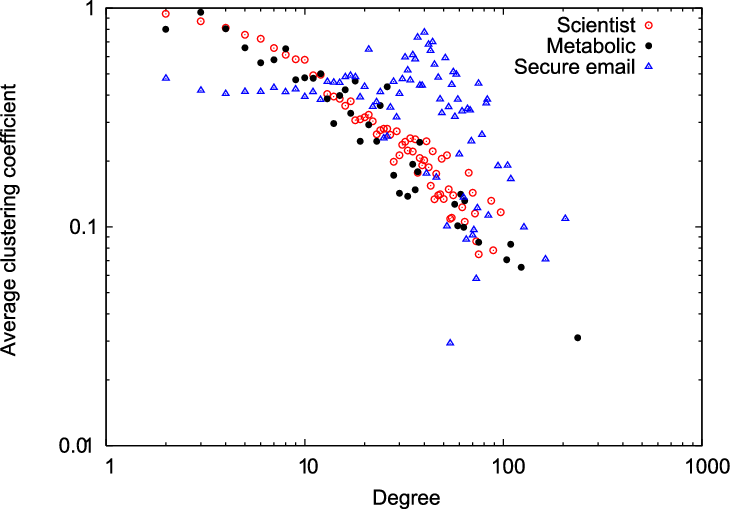} \caption{ Average clustering coefficient of
$k$-degree nodes.} \label{fig:c-k}
\end{figure}

Figure\,\ref{fig:c-k} reveals that the {\tt Scientist} network and the \texttt{Secure email} network are
fundamentally different in the relation between
clustering coefficient and node degree, yet they have similar ${\cal{R}}_{avg}$ and
${\cal{R}}_{max}$. Whereas the {\tt Scientist} network and the
\texttt{Metabolism} exhibit very similar  clustering coefficient properties, but their second-order coefficients
 are significantly different.  
 
 The above results suggest that the second-order assortative mixing is something quite
unexpected, particularly considering the work on the hierarchical organisation of complex
networks~\cite{Dorogovtsev02a,Ravasz03}.

\subsection{Common most prominent neighbour}

It is interesting to consider how often  the
most prominent contact at each end of a link is the same person, and therefore they form a triangle. Let $X$
denote the degree difference between the most prominent neighbour of the two end nodes of a link,
i.e.~$X=|K_{max}-K'_{max}|$, and $L_{<x}$ denote the number of links with $X<x$.
Table\,\ref{table:LinkCounts} shows the ratio of $L_{<4}$, $L_{<2}$ and $L_{<1}$ to the total number of
links, $L$, respectively. Note that $L_{<1}$ represents the case where $K_{max}=K'_{max}$. Also shown is
$L_{\Delta}/L$, where $L_{\Delta}$ is the number of links for which the most prominent neighbour of the two
end nodes are one and the same node and therefore forming a triangle, $L_{\Delta}\in L_{<1}$. 
Clearly the common most prominent neighbour does not provide an adequate explanation for our observations. 

\begin{table} 
	\centering
\caption{{Link ratio values of the networks under study}. }\label{table:LinkCounts}
\renewcommand{\tabcolsep}{0.5pc}
\begin{tabular}{lrrrrr}
\hline\noalign{\smallskip}
    Network             &  $^{L_{<4}}/_L$  &   $^{L_{<2}}/_L$   &     $^{L_{<1}}/_L$    &   $^{L_{\Delta}}/_L$  &   $R_{max}$  \\
\noalign{\smallskip}\hline\noalign{\smallskip}
(a) Film actor    &  34.2\%   &   34.2\%   &  34.1\%   &   34.1\% &  0.813     \\
(b) Scientist     &       50.2\%   &   45.4\%   &   42.9\%   &   41.6\%  &  0.647  \\
(c) Secure email    &       43.2\%   &   37.8\%   &   35.5\%   &   34.6\%&  0.680    \\
(d) Email         &       26.8\%   &   20.2\%   &   15.0\%   &   12.8\%    &  0.247  \\
(e) Musician     &       56.3\%   &   56.0\%   &   55.7\%   &   55.7\% &   0.307   \\
(f) Metabolism   &       51.3\%   &   51.1\%   &   50.8\%   &   50.7\% &   0.263   \\
(g) Protein        &      10.5\%   &   8.0\%    &   6.4\%    &   5.7\%   &  0.033  \\
(h) Power grid   &      68.1\%   &   38.3\%   &   19.1\%   &   7.8\%  &  0.258      \\
(i) Internet      &      20.5\%   &   20.3\%   &   20.2\%   &   20.2\%  &   0.036   \\
\noalign{\smallskip}\hline\noalign{\smallskip}
\end{tabular}
\end{table}

\subsection{Bipartite Network}

A bipartite network is a network with two non-overlapping sets of nodes $\Delta$ and $\Gamma$, where all
links must have one end node belonging to each set. For example, actors star in films, scientist write
papers, and musician play in bands. The \texttt{Film actor}, \texttt{Scientist} and \texttt{Musician}
networks under study are constructed from bipartite networks, e.g. two actors are linked if they co-star in a
film and two scientists are linked if they co-author a paper.

The \texttt{Film actor}, \texttt{Scientist} and \texttt{Musician} networks all exhibit strong second-order
assortative mixing.  It is therefore
reasonable to ask whether the second-order assortative
mixing can be attributed to the nature of bipartite networks? For example, all actors of one film constitute
a complete subgraph, in which everyone connects with the highest-degree node in the group.

However, we found no support for this hypothesis.  Firstly, the \texttt{Metabolic} network is also constructed from a bipartite network where
the two types of nodes are metabolites and reactions. Two metabolites are linked if they participate in a
reaction. The \texttt{Metabolic} network, however, does not show a
strong second-order assortative mixing.

Secondly, the \texttt{Secure email} network is a non-bipartite network, where two email users are linked by
direct email communications. It exhibits one of the strongest second-order
assortative mixing.

\begin{figure}
\centering
	\includegraphics[width=10cm]{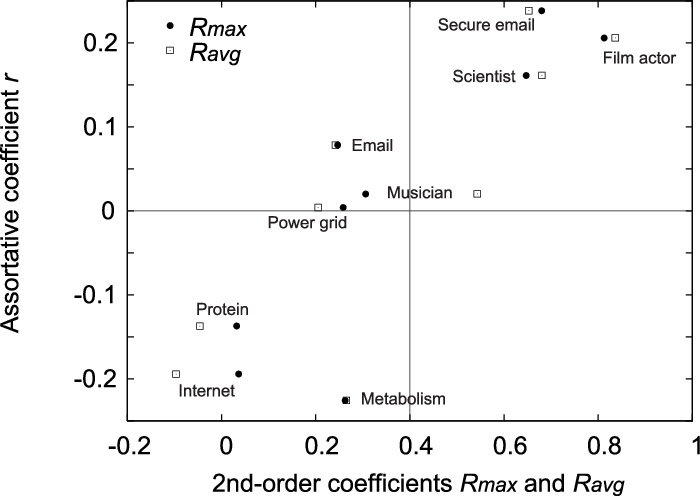} \caption{ Second-order assortative coefficients
		$R_{max}$ and $R_{avg}$ vs first-order assortative coefficient $r$.} \label{fig:r-R}
\end{figure}

\subsection{Relation Between First and Second-Order Mixing Coefficients}

Figue~\ref{fig:r-R} compares the assortative coefficient $r$ and  the second-order coefficients $R_{max}$  and $R_{avg}$ for the networks under study. 
They are seemingly loosely related.  

However, there are exceptions. Consider the {\tt Metabolic} network
and the {\tt Email} network, the former is strongly disassortative
with $r=-0.226$, whereas the later is assortative with $r=0.078$.  Yet both networks exhibit similar   values of the
second-order mixing coefficients.

\section{Conclusion}

Our experimental results demonstrated very strong-second order
assortative mixing in social networks where human are in charge of forming connections;  but weaker, or even negative
values for biological and technological networks where there is a lack of social preference.  

We examined a
larger variety of other network properties in an effort to establish
whether second-order assortative mixing was induced from other network
properties such as its power law distribution, cluster coefficient,  and bipartite
graphs.  However, although some of them might be a contributing factor, 
none of these properties was found to provide an
adequate explanation.  
We therefore conclude that second-order
assortative mixing is a new property, which reveals a new dimension to the hierarchical structure present in social networks. 

For social networks, the degree of a node is often considered a proxy for the prominence or
importance of a person. 
First order assortative mixing has then been
interpreted as indicating that if two people interact in a social
network then they are likely to have similar prominence.  
The much stronger second-order
assortative mixing suggests that there could be an even stronger social parity when measuring the prominence of a person's contacts.   
Whether our most prominent
contacts serve to introduce us  or we simply
prefer to mix with people who know similarly important people, remains an open question.

We expect that
our work will provide new clues for  studying  the structure and evolution of social networks as
well as complex networks in general.

\section*{Acknowledgments}

The authors thank Ole Winther and Sune Lehmann of DTU, Denmark for
discussions relating to the clustering coefficient.

\end{document}